\documentclass[twocolumn,aps,pre,showpacs,eqsecnum,superscriptaddress,floatfix,twoside]{revtex4}

\usepackage{graphicx}
\usepackage{graphics}
\usepackage{amsmath}
\usepackage{amssymb}
\usepackage{mathrsfs}
\usepackage{epsf}

\begin{document}

\title{Escape rate from a metastable state weakly interacting with a heat
bath driven by an external noise}

\author{Jyotipratim Ray Chaudhuri}
\email{jprc_8@yahoo.com} \affiliation{Department of Physics, Katwa
College, Katwa, Burdwan 713130, West Bengal, India}
\author{Debashis Barik}
\email{deba51@rediffmail.com} \affiliation{Indian Association for
the Cultivation of Science, Jadavpur, Kolkata 700032, India}
\author{Suman Kumar Banik}
\email{skbanik@phys.vt.edu} \altaffiliation[]{Present address:
Department of Physics, Virginia Polytechnic Institute and State
University, Blacksburg, VA 24061-0435, USA} \affiliation{Indian
Association for the Cultivation of Science, Jadavpur, Kolkata
700032, India}

\date{\today}

\begin{abstract}
Based on a system-reservoir model, where the reservoir is driven by
an external stationary, Gaussian noise with arbitrary decaying
correlation function, we study the escape rate from a metastable
state in the energy diffusion regime. For the open system we derive
the Fokker-Planck equation in the energy space and subsequently
calculate the generalized non-Markovian escape rate from a
metastable well in the energy diffusion domain. By considering the
dynamics in a model cubic potential we show that the results
obtained from numerical simulation are in good agreement with the
theoretical prediction. It has been also shown numerically that the
well known turnover feature can be restored from our model.
\end{abstract}


\maketitle


\section{Introduction}

Ever since Kramers proposed his seminal work \cite {kramers} for
chemical reaction in terms of the theory of Brownian motion in phase
space, the model and its several variants remain ubiquitous in many
areas of natural sciences. Through the years it has been a subject
of several theoretical
\cite{jth,pollak,review1,review2,semi,ralf,chaos} and experimental
\cite{expt1,expt2,lexpt} investigations for understanding the nature
of activated rate processes. In the majority of these treatments,
one is essentially concerned with a thermally equilibrated bath,
which simulates the reaction coordinate to cross the activation
energy barrier. The inherent noise of the medium is of internal
origin, which implies that the dissipative force which the system
experiences in course of its motion in the medium and the stochastic
force acting on the system as a result of the random impact from the
constituents of the medium arises from a common mechanism. From a
microscopic point of view, the system-reservoir Hamiltonian
description \cite{zwanzig,bath} suggests that the coupling of the
system and the reservoir coordinates determines both the noise and
the dissipative terms in the Langevin equation describing the motion
of the system and therefore these two entities get related through a
fluctuation-dissipation relation \cite{kubo}, which is the
characteristics of a thermodynamically closed system in contrast to
the systems driven by external noise \cite{lw,skb}. However, when
the reservoir is modulated by an external noise, it is likely that
it induces fluctuations in the polarization of the reservoir. These
fluctuations in turn may drive the system in addition to the usual
internal noise of the reservoir. Since the fluctuations of the
reservoir crucially depends on the response function; one can
envisage a connection between the dissipation of the system and the
response function of the reservoir due to external noise, from a
microscopic standpoint \cite{jrc}. At this point it is important to
mention that, a direct driving of the system coordinate breaks the
fluctuation-dissipation relation and can generate biased directed
motion that can be seen in ratchets and molecular motors
\cite{motor}. On the other hand bath modulation by an external noise
agency preserves the fluctuation dissipation relation, as a result
of which well known Kramers' turnover feature can be restored.

In many cases involving chemical systems the Markovian
representation of the Langevin equation is not valid. In the
Markovian description the time scale associated with the motion of
the thermal bath is much shorter than any relevant molecular time
scale. This assumption is practically never realized in cases where
the system coordinate is a molecular vibrational coordinate, because
the correlation time associated with the thermal bath is usually
much longer than a typical molecular vibrational period. This
observation is of no consequence for the escape rate in the strong
and moderate friction cases, where the particle is considered to be
essentially in thermal equilibrium within the well and where the
dynamics takes place only near the barrier top. Non-Markovian
effects may be important also for barrier crossing dynamics,
however, this depends on the relation between the barrier frequency
(renormalized by the presence of the friction) and the friction
coefficient. But in the low friction limit, where the well dynamics
is important, energy accumulation becomes the rate determining step.
Furthermore, reactions occurring under the nonequilibrium situation,
the well dynamics becomes crucial and become dominant in the low
friction regime. Obviously the well dynamics is governed by energy
accumulation and relaxation processes. In addition to Kramers'
treatment in the low friction, there are several treatments that
deal such situation, among which Zwanzig \cite{zwpf}, using the
assumption that reservoir is always in thermal equilibrium,
developed a procedure for reducing the classical Hamilton's
equations of motion for a one dimensional particle interacting with
a non-Markovian heat bath. The escape of a particle from a potential
well has been treated using a generalized Langevin equation in the
low friction limit by Carmeli and Nitzan (CN)\cite{cn}. Thereafter
the detailed classical analysis reveals that the rate is
significantly modified by memory effects when compared to
corresponding Kramers' theory.

While nonequilibrium, nonthermal systems have also been investigated
phenomenologically by a number of workers in several contexts
\cite{skb,rattray,nnds,masoliver,sjbe,rig1}, these treatments
concern mainly with direct driving of the system by an external
noise or a time dependent field, e.g., for examining the role of
color noise in stationary probabilities \cite{rattray}, properties
of nonlinear systems \cite{nnds}, nature of crossover
\cite{masoliver}, effect of monochromatic noise \cite{sjbe}, to
study the chemical reaction dynamics in an anisotropic solvents
\cite{rig1}. In the present paper we consider a system-reservoir
model where the reservoir is modulated by an external noise. Our
object here is to explore the role of reservoir response on the
system dynamics and to calculate the generalized escape rate from a
metastable state for a nonequilibrium open system in the energy
diffusion regime.

A number of different situations depicting the modulation of the
bath may be of physically relevant. Though the dynamics of a
Brownian particle in a uniform solvent is well-known, it is less
clear when the response of the solvent be time dependent, as in the
case of the dynamical properties of a suspension in a liquid crystal
when projected on to an anisotropic stochastic equations of motion,
or in the diffusion and reaction in supercritical liquids and growth
in living polymerization \cite{rig1,lee}. Also the space dependent
friction may be realized from the presence of a stochastic potential
in the Langevin equation \cite{rig2}. As another example, we
consider a simple unimolecular conversion from A$\rightarrow$B, say
an isomerization reaction. The reaction can be carried out in a
photochemically active solvent under the influence the external
fluctuating light intensity. Since the fluctuation in the light
intensity results in the fluctuations in the polarization of the
solvent molecules, the effective reaction field around the reactant
system gets modified \cite{nit}. In passing we mention that the
escape rate in the energy diffusion regime is just not a theoretical
issue today but has been a subject of experimental investigation
over the last two decades \cite{lexpt}.

The outlay of the paper is as follows. In section II we discuss a
system-reservoir model where the latter is modulated by an external
noise and establish an important connection between the dissipation
of the system and the response function of the reservoir due to the
external noise. The stochastic motion in energy space and
Fokker-Planck equation has been constructed in section III. We solve
the problem of energy diffusion controlled rate processes in section
IV. An explicit example with a cubic potential is worked out to
illustrate the theory in section V. The paper is concluded in
section VI.


\section{The Model: Heat bath modulated by external noise}

We consider a classical particle of mass $M$ bilinearly coupled to a
heat bath consisting of $N$ harmonic oscillators driven by an
external noise. The total Hamiltonian of such a composite system can
be written as \cite{zwanzig,bath}

\begin{equation}
\label{eq1a} H = \frac{p^2}{2M} + V(x) + \frac{1}{2} \sum_{i=1}^N
\left \{ \frac{p_i^2}{m_i} + m_i \omega_i^2 (q_i-g_ix)^2 \right \} +
H_{int} .
\end{equation}

\noindent In the above equation, $x$ and $p$ are the co-ordinate and
momentum of the system particle, respectively and $V(x)$ is the
potential energy of the system. ($q_i$,$p_i$) are the variables for
the $i$-th oscillator having frequency $\omega_i$ and mass $m_i$.
$g_i$ is the coupling constant for system-bath interaction.
$H_{int}$ is the interaction term between the heat bath and the
external noise, $\epsilon (t)$ with the following form

\begin{equation}
\label{eq2a} H_{int} = \frac{1}{2} \sum_{i=1}^N \kappa_i q_i
\epsilon (t) .
\end{equation}

\noindent The type of interaction we have considered between the
heat bath and the external noise, $H_{int}$ is commonly known as the
\textit{dipole interaction} \cite{dipole}. In Eq.(\ref{eq2a})
$\kappa_i$ denotes the strength of interaction. We consider
$\epsilon (t)$ to be a stationary, Gaussian noise processes with
zero mean and arbitrary correlation function

\begin{equation}
\label{eq3a} \langle \epsilon (t) \rangle_e = 0 \text{ and } \langle
\epsilon (t) \epsilon (t') \rangle_e = 2D\Psi (t-t')
\end{equation}

\noindent where $D$ is the external noise strength, $\Psi (t-t')$ is
the external noise kernel and $\langle \ldots \rangle_e$ implies the
averaging over the external noise processes.

Eliminating the bath degrees of freedom in the usual way (and
putting $M$ and $m_i$ equal to one) we obtain the following
generalized Langevin equation

\begin{eqnarray}
\dot{x} & = & v, \nonumber \\
\dot{v} & = & -\frac{dV}{dx} - \int_0^t dt' \gamma (t-t') v(t') +
f(t) + \pi (t) \label{eq4a}
\end{eqnarray}

\noindent where

\begin{equation}
\label{eq5a} \gamma (t) = \sum_{i=1}^N g_i^2 \omega_i^2 \cos
\omega_i t
\end{equation}

\noindent and $f(t)$ is the thermal fluctuation generated due to
system-reservoir interaction and is given by

\begin{equation}
\label{eq6a} f(t) = \sum_{i=1}^N g_i \{ [ q_i (0) - g_i x(0) ]
\omega_i^2 \cos \omega_i t + v_i (0) \omega_i \sin \omega_i t \} .
\end{equation}

\noindent in Eq.(\ref{eq4a}), $\pi (t)$ is the fluctuating force
term generated due to the external stochastic driving $\epsilon (t)$
and is given by

\begin{equation}
\label{eq7a} \pi (t) = - \int_0^t \varphi (t-t') \epsilon (t') dt' ,
\end{equation}

\noindent where

\begin{equation}
\label{eq8a} \varphi (t) = \sum_{i=1}^N g_i \kappa_i \omega_i \sin
\omega_i t .
\end{equation}

The form of Eq.(\ref{eq4a}) therefore suggests that the system is
driven by two forcing functions $f(t)$ and $\pi (t)$. The initial
conditions of the bath oscillators for a fixed choice of the initial
condition of the system degrees of freedom determines $f(t)$. To
define the statistical properties of $f(t)$, we assume that the
\textit{initial distribution} is one in which the bath is
equilibrated at $t=0$ in the presence of the system but in the
absence of the external noise agency such that $\langle
f(t)\rangle=0$ and $\langle f(t)f(t')\rangle=k_BT\gamma (t-t')$.

Now, at $t=0_+$, the external noise agency is switched on and the
bath is modulated by $\epsilon (t)$. The system is governed by
Eq.(\ref{eq4a}), where apart from the internal noise $f(t)$, another
fluctuating force $\pi (t)$ appears, that depends on the external
noise $\epsilon (t)$. Therefore, one can define an effective noise
$\xi (t) [=f(t)+\pi (t)]$ whose correlation is given by

\begin{eqnarray}
\langle \langle \xi (t) \xi (t') \rangle \rangle & = & k_BT \gamma
(t-t') + 2D \int_0^t dt'' \int_0^{t'} dt''' \varphi (t-t'')
\nonumber \\
& & \times \varphi (t'-t''') \Psi (t''-t''') \nonumber \\
& = & C(t-t') \text{ (say) },\label{eq9a}
\end{eqnarray}

\noindent along with $\langle \langle \xi (t)\rangle \rangle = 0$,
where $\langle \langle \ldots \rangle \rangle$ means we have taken
two averages independently. While writing (\ref{eq9a}) we made the
assumption $\langle \langle \xi (t) \xi (t') \rangle \rangle = C
(t-t')$ which cannot be proved unless the structure of $\varphi (t)$
is explicitly given. However, as we shall see in section V and in
the following discussion that it is a valid assumption (see,
Eqs.(2.15-2.17) and (5.5)) for a particular choice of the coupling
coefficients $g(\omega )$ and $\kappa (\omega )$ (see Eqs.(2.10) and
(2.11)) and for a stationary external noise processes (see
Eq.(5.1)). It should be emphasized that the above relation
(\ref{eq9a}) is not a fluctuation-dissipation relation due to the
appearance of the external noise intensity. Rather it serves as a
\textit{thermodynamic consistency condition}.

Let us now digress a little bit about $\pi (t)$. The statistical
properties of $\pi (t)$ are determined by the normal-mode density of
the bath frequencies, the coupling of the system with the bath, the
coupling of the bath with the external noise, and the external noise
itself. Equation (\ref{eq7a}) is the reminiscent of the familiar
linear relation between the polarization and the external field,
where $\pi$ and $\epsilon$ play the role of the former and the
latter, respectively. $\varphi (t)$ can then be interpreted as a
response function of the reservoir due to external noise $\epsilon
(t)$. The very structure of $\pi (t)$ suggests that this forcing
function, although originating from an external force, is different
from a direct driving force acting on the system. The distinction
lies at the very nature of the bath characteristics (rather than
system characteristics) as reflected in the relations (\ref{eq7a})
and (\ref{eq8a}).

In order to obtain a finite result in the continuum limit, the
coupling functions $g_i=g(\omega)$ and $\kappa_i=\kappa(\omega)$ are
chosen \cite{bravo} as $g(\omega )=g_0 / \sqrt{\tau_c}\omega$ and
$\kappa (\omega)=\kappa_0\omega\sqrt{\tau_c}$. Consequently
$\gamma(t)$ and $\varphi (t)$ reduce to the following forms:

\begin{equation}
\label{eq10a} \gamma (t) = \frac{g_0^2}{\tau_c} \int d\omega
\mathscr{D} (\omega) \cos \omega t
\end{equation}

\noindent and

\begin{equation}
\label{eq11a} \varphi (t) = g_0 \kappa_0 \int d\omega \mathscr{D}
(\omega) \omega \sin \omega t,
\end{equation}

\noindent where $g_0$ and $\kappa_0$ are constants and $1/\tau_c$ is
the cutoff frequency of the oscillator ($\tau_c$ may be
characterized as the correlation time of the bath \cite{lw} and for
$\tau_c \rightarrow 0$ we obtain $\delta$-correlated noise process).
$\mathscr{D} (\omega)$ is the density of modes of the heat bath
which is assumed to be a Lorentzian

\begin{equation}\label{eq11a1}
\mathscr{D} (\omega) = \frac{2}{\pi \tau_c (\omega^2+\tau_c^{-2})}.
\end{equation}

\noindent This assumption resembles broadly the behavior of the
hydrodynamical modes in a macroscopic system \cite{resib}. This form
of density of modes, along with the expressions of $g(\omega)$ and
$\kappa(\omega)$, allows us to write for the expression of $\varphi
(t)$ as

\begin{equation}\label{eq11a2}
\varphi (t) = (g_0 \kappa_0 /\tau_c) \exp(-t/\tau_c).
\end{equation}

\noindent From Eq.(\ref{eq10a}) and Eq.(\ref{eq11a}) one obtains
\cite{jrc}

\begin{equation}
\label{eq12a} \frac{d\gamma}{dt} = -\frac{g_0}{\kappa_0}
\frac{1}{\tau_c} \varphi (t) .
\end{equation}

\noindent Equation (\ref{eq12a}) is an important content of the
present model. This expresses how the dissipative kernel $\gamma
(t)$ depends on the response function $\varphi (t)$ of the medium
due to external noise $\epsilon (t)$ [see Eq.(\ref{eq7a})]. Such a
relation for the open system can be anticipated in view of the fact
that both the dissipation and the response function crucially depend
on the properties of the reservoir especially on its density of
modes and its coupling to the system and the external noise source.

To continue if we assume that $\epsilon (t)$ is a delta correlated
noise, \textit{i.e.}, $\langle \epsilon (t) \epsilon (t')\rangle = 2
D \delta (t-t') $, then the correlation function of $\pi(t)$ is
represented as

\begin{equation}\label{eq12a1}
\langle \pi(t) \pi(t') \rangle  = D (g_0\kappa_0)^2 \tau_c^{-1}
\exp(-|t-t'|/\tau_c)
\end{equation}

\noindent where we have neglected the transient terms ($t,t'>
\tau_c$). This equation shows how the heat bath dresses the external
noise. Though the external noise is a delta-correlated, the system
encounters it as an Ornstein-Uhlenbeck nose with the same
correlation time as the internal noise but with an intensity
depending on the couplings and the external noise strength. On the
other hand, if the external noise is an Ornstein-Uhlenbeck process
with $\langle \epsilon (t) \epsilon (t')\rangle = (D/\tau')
\exp(-|t-t'|/\tau')$ where $D$ and $\tau'$ are the strength and the
correlation time of the noise respectively, the correlation function
of $\pi(t)$ is found to be

\begin{eqnarray}\label{eq12a2}
\langle \pi(t)\pi(t') \rangle&&=\frac{(D g_0
\kappa_0)^2}{(\tau'/\tau_c)^2-1} \frac{\tau'}{\tau_c} \left\{
\frac{1}{\tau_c}\exp\left(-\frac{|t-t'|}{\tau'}\right)\right.\nonumber\\&&-\left.
\frac{1}{\tau'}\exp\left(-\frac{|t-t'|}{\tau_c}\right)\right\}
\end{eqnarray}

\noindent where we have neglected the transient terms. The dressed
external noise $\pi(t)$ now has a more complicated correlation
function with two correlation times $\tau_c$ and $\tau'$. If the
external noise-correlation time is much larger than the internal
noise correlation time, \textit{i.e.}, $\tau' \gg \tau_c$, which is
more realistic, then then the dressed noise is dominated by the
external noise, \textit{i.e.},

\begin{equation}\label{eq12a3}
\langle \pi(t)\pi(t') \rangle=\{ (D g_0 \kappa_0)^2/\tau'
\}\exp[-|t-t'|/\tau'].
\end{equation}

\noindent On the other hand, when the external noise correlation
time is smaller than the internal one, we recover Eq.(\ref{eq12a1}).


\section{Kramers equation in energy space}

To start with we first define the Fourier transform of $C(t)$ and
$\gamma (t)$ as,

\begin{eqnarray}
\label{eq1b} \widehat{C}_n (\omega) & = & \int_0^\infty dt C(t)
\exp (-in\omega t) , \\
\label{eq2b} \widehat{\gamma}_n (\omega) & = & \int_0^\infty dt
\gamma(t) \exp (-in\omega t) .
\end{eqnarray}

\noindent In the absence of external stochastic driving force
$\epsilon (t)$, the fluctuation-dissipation relation $\langle
f(t)f(t')\rangle = k_BT \gamma (t-t')$ can be expressed in the
Fourier domain as (now $C(t-t')=\langle f(t)f(t')\rangle$)

\[ \widehat{C}_n^c (\omega) = k_BT \widehat{\gamma}_n^c \]

\noindent where $\widehat{C}_n^c (\omega)$ and
$\widehat{\gamma}_n^c$ are the cosine component of $ \widehat{C}_n$
and $\widehat{\gamma}_n$, respectively. Unless the explicit form of
$\epsilon (t)$ is specified it is difficult to express the
thermodynamic consistency relation (\ref{eq9a}) in the Fourier
domain. Without loosing generality we thus will use the general form
(\ref{eq1b}-\ref{eq2b}) to derive the Fokker-Planck equation until
we use explicit form of $\epsilon (t)$. Conventionally low-friction
regime assumes the relation $\gamma \ll \omega \ll 1/\tau_c $, where
$\gamma$ is the friction arising due to interaction with the heat
bath, evaluated in the Markovian limit. $\tau_c$ is the correlation
time of the noise due to heat bath and $\omega$ is the linearized
system frequency, such relation was also considered by Kramers in
the low-friction regime as well as for the white noise case. But in
this paper we are concerned not only low-friction regime but also
with non-Markovian effect due to bath. In this context we consider
the following time scales in the dynamics relevant for energy
diffusion in the non-Markovian limit \cite{cn},

\begin{equation}
\label{eq3b} \gamma \ll 1/\tau_c \ll \omega,
\end{equation}

The separation of time scales in Eq.(\ref{eq3b}) now allow us to
write Eq.(\ref{eq4a}) into the action ($J$) and the angle ($\phi$)
co-ordinates as,

\begin{eqnarray}
\label{eq4b} \dot{J} & = & \frac{\partial x}{\partial \phi} \left
[ - \int_0^t d\tau \gamma (t-\tau) v(\tau) + \xi (t) \right ], \\
\label{eq5b} \dot{\phi} & = & \omega (J) - \frac{\partial
x}{\partial J} \left [ - \int_0^t d\tau \gamma (t-\tau) v(\tau) +
\xi (t) \right ]
\end{eqnarray}

\noindent where $v$ represents the velocity of the particle and for
the deterministic part of the system's Hamiltonian, $H=v^2/2+V(x)$
we can write

\begin{equation}
\label{eq6b} \omega (J) = \frac{dH(J)}{dJ}.
\end{equation}

\noindent Since the canonical transformation
$(x,v)\rightarrow(J,\phi)$ has been done with the deterministic part
of the Hamiltonian it is implied that $x$ and $v$ can be expanded in
terms of $J$ and $\phi$,

\begin{subequations}
\begin{eqnarray}
\label{eq7b1} x(J,\phi) & = & \sum_{n=-\infty}^\infty x_n(J)
\exp(in\phi) \\
\label{eq7b2} v(J,\phi) & = & \sum_{n=-\infty}^\infty v_n(J)
\exp(in\phi),
\end{eqnarray}
\end{subequations}

\noindent along with $x_n=x_{-n}^\ast$ and $v_n=v_{-n}^\ast$.
Differentiating Eq.(\ref{eq7b1}) with respect to time and noting
that in the action-angle variable space $\dot{\phi}=\omega(J)$ we
can write

\begin{equation}
\label{eq8b} v_n (J) = in \omega(J)x_n(J).
\end{equation}

\noindent Since we are dealing with the dynamics in the one
dimension only we can choose $J$ and $\phi$ in such a way that we
can make the simplification $x=(1/2)\sum_{n=-\infty}^\infty [x_n
\exp(in\phi) + x_n^\ast \exp(-in\phi) ]$ for $x=x^\ast$. With the
choice of phase $x=x_{-n}$ (since Im$(x_n)=0$), $x$ may be further
expressed as $x=\sum_{n=-\infty}^\infty x_n \cos n \phi$. Similarly
using Eq.(\ref{eq8b}) we get $v=\sum_{n=-\infty}^\infty v_n \sin n
\phi$ for $v_n=-v_{-n}$ (since Re$(v_n)=0$). Now inserting
Eqs.(\ref{eq7b1}-\ref{eq7b2}) in Eqs.(\ref{eq4b}-\ref{eq5b}) we
obtain

\begin{widetext}

\begin{eqnarray}
\label{eq9b} \dot{J} &=& -i \sum_{n=-\infty}^\infty
\sum_{m=-\infty}^\infty n x_n \exp (in\phi) \int_0^t d\tau \gamma
(t-\tau) v_m \exp (im\phi) + i \xi (t) \sum_{n=-\infty}^\infty n
x_n \exp (in\phi), \\
\label{eq10b} \dot{\phi} & = & \omega (J) + \sum_{n=-\infty}^\infty
\sum_{m=-\infty}^\infty \frac{\partial x_n}{\partial J} \exp
(in\phi) \int_0^t d\tau \gamma (t-\tau) v_m \exp (im\phi) - \xi(t)
\sum_{n=-\infty}^\infty \frac{\partial x_n}{\partial J} \exp
(in\phi).
\end{eqnarray}

\end{widetext}

\noindent In the equations (\ref{eq9b}) and (\ref{eq10b}), the
argument of the damping memory kernel $\gamma$ is $(t-\tau)$. Now
$\gamma$ decays to zero within the correlation time $\tau_c$. So, to
deal with the integrals of Eqs.(\ref{eq9b}-\ref{eq10b}), it is
reasonable to divide the range of integration into two parts: (a)
$|t-\tau| \leqslant \tau_c$ and (b) $t \gg \tau_c$. Thus following
CN \cite{cn} we can write

\[ \phi (t) = \phi [\tau+(t-\tau)] \simeq \phi (\tau) +
\left. \frac{\partial \phi}{\partial t}\right |_{t=\tau} (t-\tau),
\]

\noindent neglecting higher terms of $\tau_c$. It follows that

\begin{equation}
\label{eq11b} \phi (\tau) \simeq  \phi (t) - (t-\tau) \omega \text{
and } v_m (\tau)  \simeq v_m (t).
\end{equation}

\noindent Equation (\ref{eq11b}) is reasonable approximation so far
as the of Eqs.(\ref{eq9b}-\ref{eq10b}) are concerned. Within the
integral, we therefore manipulate the behavior of $\phi$ and $v_m$
for a time upto which $\gamma (t-\tau)$ exists and also for the
observational time at which $\gamma$ has decayed to zero. So, more
specifically we can write for $|t-\tau| \leqslant \tau_c$,

\begin{eqnarray}
& & \int_0^t d\tau \gamma (t-\tau) v_m (\tau) \exp[im\phi(\tau)]
\nonumber \\
& & \simeq v_m (t) \exp[im\phi (t)] \int_0^t d\tau \gamma
(t-\tau) \exp[-im(t-\tau)\omega] \nonumber \\
& & \label{eq12b}
\end{eqnarray}

\noindent and for $t \gg \tau_c$, using Eq.(\ref{eq2b}) we have

\begin{eqnarray}
&& \int_0^t d\tau \gamma (t-\tau) v_m (\tau) \exp[im\phi (\tau)]
\nonumber \\
&& \simeq v_m (t) \exp[im\phi (t)] \widehat{\gamma}_m (\omega).
\label{eq13b}
\end{eqnarray}

\noindent Putting Eq.(\ref{eq13b}) which takes into account the
observational time scale, in Eqs.(\ref{eq9b}) and (\ref{eq10b}) we
get

\begin{eqnarray}
\dot{J} &=& -i \sum_{n=-\infty}^\infty \sum_{m=-\infty}^\infty n x_n
v_m \widehat{\gamma}_m (\omega) \exp
[i(n+m)\phi] \nonumber \\
& & + i \xi (t) \sum_{n=-\infty}^\infty n
x_n \exp (in\phi), \label{eq14b} \\
\dot{\phi} & = & \omega (J) + \sum_{n=-\infty}^\infty
\sum_{m=-\infty}^\infty \frac{\partial x_n}{\partial J} v_m
\widehat{\gamma}_m (\omega) \exp
[i(n+m)\phi] \nonumber \\
& & - \xi(t) \sum_{n=-\infty}^\infty \frac{\partial x_n}{\partial J}
\exp (in\phi). \label{eq15b}
\end{eqnarray}

\noindent Now we are in a position to formulate the Fokker-Planck
equation. For this we follow the method proposed by CN \cite{cn}
which is based on Kramers-Moyal expansion of the transition
probability that connects the probability distribution function
$P(J,\phi,t)$ at time $t$ with that of $P(J,\phi,t+\tau)$ at a later
time $t+\tau$ for small $\tau$, given that we know the moments of
the distribution. The time evolution of the probability distribution
$P(J,\phi,t)$ is determined by the equation,

\begin{eqnarray}
\frac{\partial P}{\partial t} & = & \lim_{\tau \rightarrow 0+} \left
[ \frac{1}{\tau} \sum_{n=1}^\infty \frac{(-1)^n}{n!}
\sum_{m,k=0;(m+k=n)} \left (
\frac{\partial}{\partial J} \right )^m \right. \nonumber \\
& & \left. \times \left ( \frac{\partial}{\partial \phi} \right )^k
\{ \langle \langle (\Delta J_t)^m (\Delta \phi_t)^k \rangle \rangle
P \} \right ], \label{eq16b}
\end{eqnarray}

\noindent where $\Delta J_t = \Delta J_t (\tau) = J(t+\tau)-J(t)$
and $\Delta \phi_t = \Delta \phi_t (\tau) =
\phi(t+\tau)-\phi(\tau)$. At this juncture it is worth recalling
that $\tau$ is the coarse-grained time scale over which the
probability distribution function evolves, whereas $\tau_c$ is the
correlation time, which due to low damping is much smaller than
$\tau$. The low value of $\gamma$ prompts us to take $1/\gamma$ as
the largest time scale for the entire problem. However, the
reciprocal of the frequency of oscillation, i.e., $1/\omega$, is the
smallest time scale. Our task is now to calculate the moments of the
form $\langle \langle (\Delta J_t)^m (\Delta \phi_t)^k \rangle
\rangle$ where $\langle\langle \cdots \rangle\rangle$ means that we
have taken the two averages independently.

To evaluate the moments we make use of the following standard
procedure \cite{cn,lax}

\begin{eqnarray}
\label{eq17b} \Delta J_t (\tau) & = & \int_0^\tau ds \dot{J}
[J(t+s),\phi (t+s),t+s] \\
\label{eq18b} \Delta \phi_t (\tau) & = & \int_0^\tau ds \dot{\phi}
[J(t+s),\phi (t+s),t+s]
\end{eqnarray}

\noindent where $\dot{J}$ and $\dot{\phi}$ are given by
Eqs.(\ref{eq14b}) and (\ref{eq15b}), respectively.

The non-Markovian nature, i.e., $\tau_c$ is finite but $\tau_c <
\tau$, of the present problem allow us to consider all orders of
$\tau$ in Eq.(\ref{eq16b}). But, since $\partial P/\partial t$ is
evaluated in the limit $\tau \rightarrow 0_+$, terms linear in
$\tau$ are taken while all the higher powers are neglected. We then
recast Eqs.(\ref{eq14b}) and (\ref{eq15b}) in the following form

\begin{eqnarray}
\dot{J} & = & -\sum_{n=-\infty}^{\infty} \sum_{m=-\infty}^{\infty}
B_{nm} (J) \exp [i(n+m)\phi] \nonumber
\\
& & + \xi (t) \sum_{n=-\infty}^{\infty} \sigma_n (J) \exp
(in\phi) \label{eq19b} \\
\dot{\phi} & = & \omega (J) + \sum_{n=-\infty}^{\infty}
\sum_{m=-\infty}^{\infty} C_{nm} (J) \exp [i(n+m)\phi] \nonumber
\\
& & - \xi (t) \sum_{n=-\infty}^{\infty} \mu_n (J) \exp(in\phi)
\label{eq20b}
\end{eqnarray}

\noindent where

\begin{eqnarray}
\label{eq21b} \sigma_n (J) & = & inx_n (J), \\
\label{eq22b} \mu_n (J) & = & \frac{dx_n(J)}{dJ}, \\
\label{eq23b} B_{nm}(J) & = & inx_n(J)v_m(J)\widehat{\gamma}_m [\omega(J)], \\
\label{eq24b} C_{nm} (J) & = & \left [ \frac{dx_n(J)}{dJ} \right ]
v_m (J) \widehat{\gamma}_m [\omega(J)].
\end{eqnarray}

Finally the moments can be calculated using the standard iterative
process prescribed by CN \cite{cn} and they are of the following
form

\begin{eqnarray}
\langle \langle \Delta J_t (\tau) \rangle \rangle & = & - 2 \tau
\sum_{n=1}^\infty n^2 \left [ \omega |x_n|^2 \widehat{\gamma}_n^c
(\omega) \right. \nonumber \\
& & \left. - \frac{d}{dJ} \{ |x_n|^2 \widehat{C}_n^c (\omega) \}
\right ], \label{eq25b} \\
\langle \langle \Delta \phi_t (\tau) \rangle \rangle & = & \omega
\tau + \tau \sum_{n=1}^\infty n \left [ \omega \widehat{\gamma}_n^s
(\omega) \frac{d |x_n|^2}{dJ} \right.
\nonumber \\
\label{eq26b} & & \left. - \frac{d}{dJ} \left ( \widehat{C}_n^s
(\omega) \frac{d |x_n|^2}{dJ} \right ) \right ],
\end{eqnarray}

\begin{eqnarray}
\label{eq27b} & & \langle \langle [\Delta J_t (\tau)]^2\rangle
\rangle = 4 \tau \sum_{n=1}^\infty n^2 |x_n|^2
\widehat{C}_n^c (\omega), \\
\label{eq28b} & & \langle \langle [\Delta \phi_t (\tau)]^2\rangle
\rangle = 4 \tau \sum_{n=1}^\infty n^2 \left | \frac{d x_n }{dJ}
\right |^2 \widehat{C}_n^c (\omega), \\
\label{eq29b} & & \langle \langle \Delta J_t (\tau) \Delta \phi_t
(\tau) \rangle \rangle = 0,
\end{eqnarray}

\noindent where

\begin{subequations}
\begin{eqnarray}
\widehat{\gamma}_n^c & = & \int_0^\infty dt \gamma (t) \cos
(n\omega t), \\
\widehat{\gamma}_n^s & = & \int_0^\infty dt \gamma (t) \sin
(n\omega t), \\
\widehat{C}_n^c & = & \int_0^\infty dt C(t) \cos (n\omega t), \\
\widehat{C}_n^s & = & \int_0^\infty dt C(t) \sin (n\omega t).
\end{eqnarray}

\noindent Also

\begin{eqnarray}
\widehat{\gamma}_n (\omega) & = & \widehat{\gamma}_n^c (\omega) -
i \widehat{\gamma}_n^s (\omega), \\
\widehat{C}_n (\omega) & = & \widehat{C}_n^c (\omega) - i
\widehat{C}_n^s (\omega).
\end{eqnarray}

\end{subequations}

\noindent In the absence of the external noise $\epsilon (t)$,
$\widehat{C}_n (\omega)$ reduces to $\widehat{C}_n (\omega) = k_BT
\widehat{\gamma}_n (\omega)$ for which
Eqs.(\ref{eq25b})-(\ref{eq29b}) becomes \cite{cn}

\begin{eqnarray*}
& & \langle \Delta J_t (\tau) \rangle  = -2\tau \sum_{n=1}^\infty
n^2 \left ( \omega -k_BT \frac{d}{dJ} \right ) (|x_n|^2
\widehat{\gamma}_n^c ), \nonumber \\
& & \langle \Delta \phi_t (\tau) \rangle = \omega \tau + \tau
\sum_{n=1}^\infty n \left ( \omega -k_BT \frac{d}{dJ} \right )
\left ( \frac{d |x_n|^2}{dJ} \widehat{\gamma}_n^s \right ), \nonumber \\
& & \langle [\Delta J_t (\tau)]^2\rangle = 4 \tau k_BT
\sum_{n=1}^\infty n^2 |x_n|^2
\widehat{\gamma}_n^c , \nonumber \\
& & \langle [\Delta \phi_t (\tau)]^2\rangle = 4 \tau k_BT
\sum_{n=1}^\infty n^2 \left | \frac{d x_n }{dJ} \right |^2
\widehat{\gamma}_n^c , \nonumber \\
& & \langle \Delta J_t (\tau) \Delta \phi_t (\tau) \rangle = 0.
\end{eqnarray*}

\noindent Note that in the above unnumbered equations there is only
one averaging, $\langle\cdots\rangle$ instead of two averaging,
$\langle \langle \cdots \rangle \rangle$ used in this article. This
is due to the fact that in the present model we make an extra
averaging over the external noise processes in addition to the usual
thermal averaging procedure.

Inserting Eqs.(\ref{eq25b})-(\ref{eq29b}) in Eq.(\ref{eq16b}) and
neglecting terms with $n>2$ we obtain the Fokker-Planck equation for
$P(J,\phi,t)$ as

\begin{eqnarray}
\frac{\partial P(J,\phi,t)}{\partial t} & = &
\frac{\partial}{\partial J} \left [ \varepsilon (J) \left \{
\frac{\widehat{C}_n^c (\omega)}{\widehat{\gamma}_n^c (\omega)}
\frac{\partial}{\partial J} + \omega (J) \right \} P \right ]
\nonumber \\
& & + \Gamma (J) \frac{\partial^P}{\partial \phi^2} - \Omega (J)
\frac{\partial P}{\partial \phi}, \label{eq30b}
\end{eqnarray}

\noindent where
\begin{eqnarray}
& & \label{eq31b} \varepsilon (J) = 2 \sum_{n=1}^\infty n^2 |x_n|^2
\widehat{\gamma}_n^c (\omega), \\
& & \label{eq32b} \Gamma (J) = 2 \sum_{n=1}^\infty n^2 \left |
\frac{d x_n}{dJ} \right |^2 \widehat{C}_n^c (\omega), \\
& & \Omega (J) = \omega + \sum_{n=1}^\infty n \left [ \omega
\widehat{\gamma}_n^s \frac{d |x_n|^2}{dJ} - \frac{d}{dJ} \left (
\widehat{C}_n^s \frac{d |x_n|^2}{dJ} \right ) \right ].
\nonumber \\
& & \label{eq33b}
\end{eqnarray}

\noindent For a distribution function that is initially ($t=0$)
independent of $\phi$ the diffusion equation in action space becomes

\begin{equation}
\label{eq34b} \frac{\partial P(J,t)}{\partial t} =
\frac{\partial}{\partial J} \left [ \varepsilon (J) \left \{ \Lambda
\frac{\partial}{\partial J} + \omega (J) \right \} P \right ],
\end{equation}

\noindent where

\begin{equation}
\label{eq35b} \Lambda = \Lambda (\omega_0) \simeq
\frac{\widehat{C}_n^c(\omega_0)}{\widehat{\gamma}_n^c (\omega_0)}.
\end{equation}

\noindent Here $\omega_0$ is the linearized frequency and $\Lambda$
plays the typical role of $k_BT$ which for $\epsilon (t)=0$ becomes
equal to $k_BT$. Now by virtue of Eq.(\ref{eq6b}) $\omega
(J)=\partial H/\partial J=dE/dJ$. Expressing

\begin{equation}
\label{eq36b} \omega (J) = \nu (E),
\end{equation}

\noindent we have

\begin{equation}
\label{eq37b} \frac{\partial}{\partial J} = \nu (E)
\frac{\partial}{\partial E}.
\end{equation}

\noindent With this transformation, for an external noise driven
bath, the Kramers equation for energy diffusion [Eq.(\ref{eq34b})]
becomes

\begin{eqnarray}
\frac{\partial P(E,t)}{\partial t} = \frac{\partial}{\partial E}
\left [ D(E) \left ( \frac{\partial}{\partial E} + \frac{1}{\Lambda}
\right ) \nu (E)
P (E,t) \right ], \nonumber \\
\label{eq38b}
\end{eqnarray}

\noindent with the following diffusion coefficient

\begin{equation}
D(E) = \nu (E) 2 \Lambda (\omega_0) \sum_{n=1}^\infty n^2 |x_n|^2
\int_0^\infty dt \gamma (t) \cos [n \nu (E) t]. \label{eq39b}
\end{equation}

\noindent \textit{Eq.(\ref{eq38b}) is the first key result of the
present article}. The equation is valid for arbitrary temperature
and noise correlation. The prime quantities that determine the
equation for energy diffusion (\ref{eq38b}) are the diffusion
coefficient $D$; the open system analogue of $k_BT$, $\Lambda$; and
the frequency of the dynamical system, $\nu (E)$. Although the
expression for diffusion coefficient (\ref{eq39b}) looks bit
complicated and formal due to the appearance of the Fourier
coefficients $x_n$ in the summation, it is possible to read the
various terms in $D(E)$ in the following way. $D(E)$ is essentially
an approximate product of three terms, $\Lambda (\omega_0)$,
$\int_0^\infty dt \gamma (t) \cos [n \nu (E) t]$, and $\nu (E)
\sum_{n=1}^\infty n^2 |x_n|^2$, where the $n$ dependence of the
latter two contributions have been separated out for interpretation.
The first term is an analogue of $k_BT$ for the open system, the
integral is the Fourier transform of the memory kernel, while the
sum can be shown to be equal to $J$ (see Appendix D of CN
\cite{cn}), the action variable. For a system only coupled to a heat
bath, i.e., for no external driving, $D(E)$ reduces to the
expression derived by CN \cite{cn}.


\section{Energy diffusion controlled rate of escape}

The classical treatment of memory effects in the energy diffusion
controlled escape is now well documented in the literature
\cite{cn,hw,gh}. To address the corresponding problem for the open
system we first rewrite the Kramers equation (\ref{eq38b}) in the
form of a continuity equation

\begin{equation}
\label{eq1c} \frac{\partial P(E,t)}{\partial t} + \frac{\partial
j_E}{\partial E} = 0,
\end{equation}

\noindent where $j_E$ is the stationary flux along the energy
coordinate and is given by

\begin{equation}
\label{eq2c} j_E = - D(E) \left [ \frac{\partial}{\partial E} +
\frac{1}{\Lambda} \right ] \nu (E) P_{st} (E),
\end{equation}

\noindent with $P_{st}$ being the stationary probability
distribution. For zero current condition, we have the stationary
distribution, $p_{st}$ at the source well

\begin{equation}
\label{eq3c} p_{st} (E) = \frac{N^{-1}}{\nu(E)} \exp(-E/\Lambda)
\end{equation}

\noindent where $N$ is the normalization constant. Here it is
important to mention that for $\epsilon (t)=0$, one has
$p_{st}=P_{eq}$. We now define the rate of escape $k$ as flux over
population \cite{farkas}

\begin{equation}
\label{eq4c} k = j_E/n_a
\end{equation}

\noindent where $n_a$ is the total population at the source well,

\begin{equation}
\label{eq5c} n_a = \int_0^{E_b} P(E)dE.
\end{equation}

\noindent Here $E_b$ is the value of the activation barrier.
Following B\"uttiker, Harris, and Landauer (BHL) \cite{bhl} we use a
Kramers like ansatz

\begin{equation}
\label{eq6c} P(E) = \eta (E) p_{st} (E)
\end{equation}

\noindent to arrive at

\begin{equation}
\label{eq7c} j_E = - D(E) \nu (E) p_{st} (E) \frac{\partial \eta (E)
}{\partial E}.
\end{equation}

\noindent Integrating the above expression from $E=E_1\simeq\Lambda$
to $E=E_b$, one derives an expression for energy independent current
$j_E$ (with $E \leqslant E_b$) as

\begin{eqnarray}
j_E & = & \frac{\eta (\Lambda)-\eta (E_b)}{\int_\Lambda^{E_b}
\frac{dE}{D(E)\nu(E)p_{st}(E)}} \nonumber \\
& = & [1-\eta (E_b)] D(E_b) \frac{N^{-1}}{\Lambda}
\exp(-E_b/\Lambda), \label{eq8c}
\end{eqnarray}

\noindent where we have used the boundary condition $\eta (\Lambda)
\simeq 1$.

Following the original reasoning by BHL we now allow an outflow
$j_{out}$ from each energy range $E$ to $E+dE$, with each $E$
satisfying the condition $E\geqslant E_b$. Then we can write

\begin{equation}
\label{eq9c} dj_{out} = \alpha \nu (E) \eta (E) p_{st} (E) dE,
\end{equation}

\noindent which is compensated by a divergence in the vertical flow

\begin{equation}
\label{eq10c} \frac{dj_E}{dE} = \alpha \nu (E) \eta (E) p_{st} (E).
\end{equation}

\noindent Here $\alpha$ is a parameter that has been set
approximately equal to one by BHL, though in general the parameter
$\alpha$ is not always equal to one \cite{review2,alpha}. Inserting
the expression for nonequilibrium current (Eq.(\ref{eq7c})) in the
above expression we obtain an ordinary differential equation for
$\eta (E)$

\begin{equation}
\label{eq11c} D (E) \frac{d^2\eta}{dE^2} + \left [
\frac{dD}{dE}-\frac{D(E)}{\Lambda} \right ] \frac{d\eta}{dE} -\alpha
\eta (E) = 0.
\end{equation}

\noindent Within small energy range above $E_b$ one can assume
essentially a constant diffusion coefficient, i.e.,
$dD(E)/dE|_{E\simeq E_b} = 0$ for $E \geqslant E_b$. Now
substituting a trial solution of the form $\eta (E) = \mathscr{C}
\exp (sE/\Lambda) $ for $s<0$, in Eq.(\ref{eq11c}) we have

\begin{equation}
\label{eq13c} s_- = \frac{1}{2} \left [ \left ( 1 + \frac{4\alpha
\Lambda^2}{D(E_b)} \right )^{1/2} - 1 \right ].
\end{equation}

\noindent Setting $\eta (E) = \eta (E_b) \exp[s(E-E_b)/\Lambda]$ and
putting this into Eq.(\ref{eq7c}) and comparing this with the right
hand side of Eq.(\ref{eq8c}) we have

\begin{equation}
\label{eq14c} \eta (E_b) = 1/(1-s) \text{ for } s<0.
\end{equation}

\noindent Thus escape rate $k$ can be written as

\begin{equation}
\label{eq15c} k = j_E \left [ \int_0^{E_b} \eta (E) p_{st} (E) dE
\right ]^{-1}.
\end{equation}

\noindent Making use of Eq.(\ref{eq14c}) in Eq.(\ref{eq8c}) and the
resulting expression for $j_E$  in Eq.(\ref{eq15c}) we obtain

\begin{equation}
\label{eq16c} k = \frac{-s}{1-s} \left [ \frac{\int_0^{E_b} \eta (E)
p_{st} (E) dE}{(N^{-1}/\Lambda) D(E_b) \exp (-E_b/\Lambda)} \right
]^{-1}.
\end{equation}

\noindent For the dynamics at the bottom we have $\eta \rightarrow
1$. For $\omega_0$ being the frequency at the bottom of the source
well we now calculate the total population of the source well,

\begin{eqnarray}
n_a & = & \int_{-\infty}^\infty \int_{-\infty}^\infty p_{st} (E)
dx dp \nonumber \\
& = & N^{-1} (2\pi \Lambda/\omega_0). \label{eq17c}
\end{eqnarray}

\noindent So, for the external noise driven heat bath the
non-Markovian rate of escape from a metastable well in the low
friction regime is given by

\begin{eqnarray}
k & = & \left [ \frac{ \{ 1 + (4 \alpha \Lambda^2/D(E_b)) \}^{1/2} -
1 }{ \{ 1 + (4 \alpha \Lambda^2/D(E_b)) \}^{1/2} + 1 }
\right ] \frac{D(E_b)}{\Lambda^2} \nonumber \\
& & \times \omega_0 \exp (-E_b/\Lambda). \label{eq18c}
\end{eqnarray}

\noindent \textit{Eq.(\ref{eq18c}) is the second key result of
present paper}.


\section{Specific example: Heat bath driven by external color
noise}

As a specific example, we consider that the heat bath is modulated
externally by a colored noise $\epsilon(t)$ with noise correlation

\begin{equation}
\label{eq1d} \langle \epsilon(t) \epsilon(t') \rangle =
\frac{D_e}{\tau_e} \exp \left[- \frac{|t-t'|}{\tau_e}\right]
\end{equation}

\noindent where $D_e$ and $\tau_e$ are the strength and the
correlation time of the external noise, respectively. In addition to
that we also consider the internal noise $f(t)$ to be white. The
effective Gaussian Ornstein-Uhlenbeck noise $\xi(t)=f(t)+\pi(t)$
will have an intensity $D_R$ and a correlation time $\tau_R$ given
by \cite{lw}

\begin{eqnarray}
D_R &=& \int_0^\infty \langle \xi(t)\xi(0) \rangle dt,
\label{eq1d1}\\
\tau_R &=& \frac{1}{D_R}\int_0^\infty t \langle \xi(t)\xi(0) \rangle
dt. \label{eq1d2}
\end{eqnarray}

\noindent Following the above definitions and using
Eq.(\ref{eq12a3}) we have

\begin{equation}
\label{eq3d} D_R=g_0^2(k_B T+D_e \kappa_0^2) \text{ and } \tau_R =
\frac{D_e g_0^2 \kappa_0^2}{D_R} \tau_e.
\end{equation}

\noindent It is important to mention here that since we are treating
the internal noise processes to be a delta correlated one($\tau_c
\rightarrow 0$), $\tau_c$ does not appear explicitly in the
expression of $D_R$ and $\tau_R$. With this the effective noise
$\xi(t)$ becomes a colored noise and its correlation is given by

\begin{equation}
\label{eq2d} \langle\langle \xi (t) \xi (t')
\rangle\rangle=\frac{D_R}{\tau_R} \exp\left[
-\frac{|t-t'|}{\tau_R}\right].
\end{equation}

To study the dynamics we consider a model cubic potential of the
form $V(x)=A x^2 - B x^3$ where $A$ and $B$ are two constant
parameters with $A > 0$ and $B > 0$. The diffusion coefficient
$D(E_b)$ in the internal white noise limit reduces to

\begin{equation}
\label{eq4d} D(E_b)=g_0^2 \Lambda(\omega_0) J
\end{equation}

\noindent where the action, $J$ is represented as \cite{cn}

\begin{equation}
\label{eq5d} J=2 \nu(E_b) \sum_{n=1}^\infty n^2 |x_n|^2.
\end{equation}

\noindent and can be calculated using the following standard form

\begin{equation}
\label{eq6d} J = \frac{1}{\pi} \int_{x_1}^{x_2} v \; dx
\end{equation}

\noindent where $x_1$ and $x_2$ are the two turning points of
oscillation for which $v$ is equal to zero and they both corresponds
to total system energy $E$. In principle, they are the first two
roots (in ascending order of magnitude) of the cubic equation

\begin{equation}
\label{eq7d} -A x^2 + B x^3 + E =0.
\end{equation}

\noindent For an external color noise driven heat bath
$\Lambda(\omega_0)$ (see Eq.(\ref{eq35b})) reduces to

\begin{equation}
\label{eq8d} \Lambda(\omega_0)=\frac{k_B T + D_e \kappa_0^2}{
1+\omega_0^2 \tau_e^2} .
\end{equation}

We then numerically solve the Langevin equation (\ref{eq4a}) using
the second order stochastic Heun algorithm \cite{heun1,heun2}. To
ensure the stability of our simulation we have used a small time
step $\Delta t=0.001$  with $\Delta t/\tau_R \ll 1$. The numerical
rate has been defined as the inverse of the mean first passage time
\cite{mfpt,db}. The mean first passage time have been calculated by
averaging over 5000 trajectories. The value of the other parameters
used are given in the caption of Figs.(1) and (2).

In his dynamical theory of chemical reactions Kramers identified two
distinct regimes of stationary nonequilibrium states in terms of
dissipation constant ($\gamma$). The essential result of Kramers'
theory is that the rate varies linearly in weak dissipation regime
(characterized of diffusion of energy) and inversely in the
intermediate to strong damping regime (spatial diffusion limited
regime). That is, in between the two regimes the rate constant as a
function of dissipation constant exhibits a bell-shaped curve known
as Kramers' turnover \cite{review1,review2}. In the traditional
system reservoir model the dissipation and the fluctuation, both
originating from a common source, the reservoir, are connected
through the fluctuation-dissipation relation. A typical signature of
this relation can be seen through the turnover phenomenon in
Kramers' dynamics. Whereas for a thermodynamic open system where the
heat bath is modulated by an external noise, both the dissipation
and the response function depend on the properties of the reservoir,
mainly on its density of modes and its coupling to the system and
the external noise source. By virtue of this connection between the
dissipation and the external noise source, Eq.(\ref{eq9a}) plays the
typical role of the thermodynamic consistency relation, an analogue
of the fluctuation-dissipation relation for thermodynamic closed
system, for which one can expect turnover like feature in Kramers'
dynamics (for the open system). So, for the external color noise
driven bath we first wanted to check whether Kramers' turnover
feature can be restored from our model. In Fig.~1  we have plotted
the rate constant, $k$ obtained from Langevin simulation, for a wide
range of damping constant, $g_0^2$ for different values of external
noise correlation time, $\tau_e$. The figure shows usual Kramers'
turnover of the rate constant with variation of the damping
constant. The shift of the maxima occurs as the external noise
correlation time varies, a typical effect of the bath modulation.

Next we compared the theoretical result (\ref{eq18c}) with the
numerical simulation data.  In Fig.~2 we have plotted the rate
constant, $k$ against the damping constant, $g_0^2$ in the weak
damping domain ($0.001 \leqslant g_0^2 \leqslant 0.01 $) for
different values of the external noise correlation time, $\tau_e$.
What we observe is that the agreement between the theoretical
prediction and numerical simulation is quite satisfactory.

\begin{figure}[!t]
\includegraphics[width=0.7\linewidth,angle=-90]{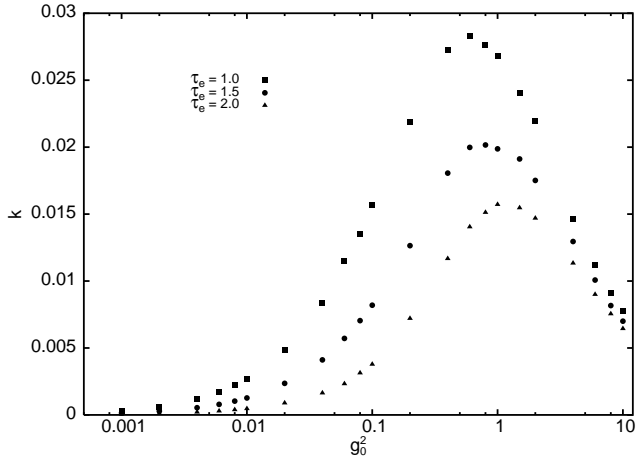}
\caption{\label{fig2} Turnover phenomenon for external color noise
driven bath. Parameters used are $k_BT = 0.1$, $D_e = 1.0$,
$\kappa_0^2 = 5.0$, $\alpha = 1.0$, $A = 0.5$ and $E_b = 5.0$ (scale
arbitrary).}
\end{figure}

\begin{figure}[!t]
\includegraphics[width=0.7\linewidth,angle=-90]{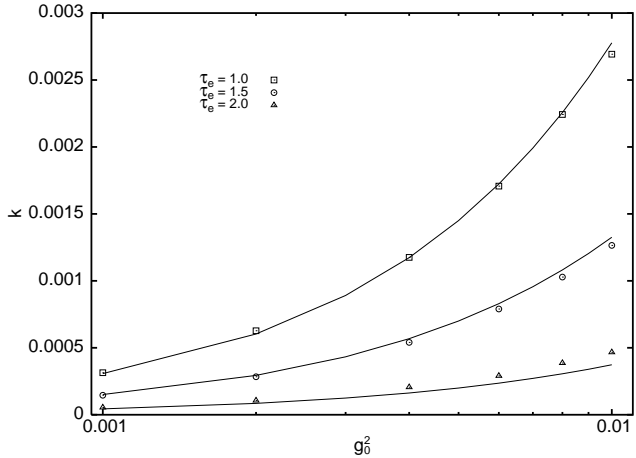}
\caption{\label{fig1} Barrier crossing rate in the low friction
regime ($0.001 \leqslant g_0^2 \leqslant 0.01$), a comparison
between theoretical prediction, Eq.\ref{eq18c} (solid lines) and
Langevin simulation. Parameters used are $k_BT = 0.1$, $D_e=1.0$,
$\kappa_0^2 = 5.0$, $\alpha = 1.0$, $A = 0.5$ and $E_b = 5.0$ (scale
arbitrary).}
\end{figure}


\section{Conclusion}

Based on a simple system-reservoir Hamiltonian approach, we have
studied the behavior of a subsystem coupled to a heat bath where the
heat bath is modulated by an external stationary, Gaussian noise
processes with arbitrary decaying correlation function, thereby
making the system thermodynamically open. For such an open system we
have analytically derived the generalized steady state Kramers'
escape rate from a metastable well in the low friction regime. The
main conclusions of the present work are the following:

(i) Since the reservoir is driven by the external noise and the
dissipative properties of the system depend on the reservoir, we
have established a simple relation between the dissipation and the
response function of the medium due to external noise. This relation
is important for identifying the effective temperature of the heat
bath characterizing the stationary state of the thermodynamically
open system.

(ii) We then followed the dynamics of the open system in the energy
space and derived the corresponding Fokker-Planck equation with
diffusion coefficient containing the effective temperature like
quantity which is an open system analogue of $k_BT$. Following the
standard approach we then derived the generalized non-Markovian
Kramers' escape rate from a metastable well in the energy diffusion
regime.

(iii) From the point of view of the realistic situation we
considered the special case where the internal noise is white and
the external noise is colored and have calculated the escape rate
for a model cubic potential. We have shown that the theoretical
prediction agrees reasonably well with numerical simulation. In
addition to that we have also shown that our model recovers the
turnover feature of the Kramers' dynamics when the external noise
modulates the reservoir.

(iv) Finally, as shown in the present work one can easily tune the
external noise parameters from outside which can be used to study
the effect of several kinds of noise properties, e.g., long tail
gaussian noise \cite{ralf}, in Kramers' dynamics. Another suitable
candidate for studying the escape rate dynamics can be irreversibly
driven environments \cite{jrc2,rig3}. In our future communication we
would like to pursue such theoretical analysis.


\begin{acknowledgments}
The authors wishes to thank Professor Deb Shankar Ray for critical
comments and suggestions. This work was partially supported by the
Department of Science and Technology, Government of India.
SKB acknowledges support from Virginia Tech through ASPIRES award
program.
\end{acknowledgments}



\begin{thebibliography}{99}

\bibitem{kramers} H.A. Kramers, Physica (Amsterdam) \textbf{7}, 284
(1940).

\bibitem{jth} R.F. Grote and J.T. Hynes, J. Chem. Phys. \textbf{73}, 2715
(1980).

\bibitem{pollak} E. Pollak, J. Chem. Phys. \textbf{85}, 865 (1986).

\bibitem{review1} P. H\"anggi, P. Talkner, and M. Borkovec, Rev.
Mod. Phys. \textbf{62}, 251 (1990).

\bibitem{review2} V.I. Mel'nikov, Phys. Rep. \textbf{209}, 1
(1991).

\bibitem{semi} J. Ray Chaudhuri, B.C. Bag, and D.S. Ray, J. Chem.
Phys. \textbf{111}, 10852 (1999).

\bibitem{ralf} R. Metzler and J. Klafter, Chem. Phys. Lett.
\textbf{321}, 238 (2000); A.V. Chechkin, V.Yu. Gonchar, J. Klafter,
and R. Metzler, Europhys. Lett. \textbf{72}, 348 (2005).

\bibitem{chaos} E. Pollak and P. Talkner, Chaos \textbf{15},
026116 (2005).

\bibitem{expt1} E.W.-G. Diau, J.L. Herek, Z.H. Kim, and A.H.
Zewail, Science \textbf{279}, 847 (1998).

\bibitem{expt2} L.I. McCann, M. Dykman, and B. Golding, Nature
\textbf{402}, 785 (1999); J. Hales, A. Zhukov, R. Roy, and M.I.
Dykman, Phys. Rev. Lett. \textbf{85}, 78 (2000).

\bibitem{lexpt} K. Luther, J. Schroeder, J. Troe, and U.
Unterberg, J. Phys. Chem. \textbf{84}, 3072 (1980); B. Otto, J.
Schroeder, and J. Troe, J Chem. Phys. \textbf{81}, 202 (1984); K.
Hara, N. Ito, and O. Kajimoto, J. Chem. Phys. \textbf{110}, 1662
(1999).

\bibitem{zwanzig} R. Zwanzig, J. Stat. Phys. \textbf{9}, 215
(1973); K. Lindenberg and V. Seshadri, Physica A \textbf{109}, 483
(1981).

\bibitem{bath} M.I. Dykman and M.A. Krivoglaz,
Phys. Stat. Sol. (b) \textbf{48}, 497 (1971); \textit{ibid}, in
\textit{Soviet Physics Reviews}, edited by I.M. Khalatnikov, Vol.~5
(Harwood, New York, 1984) pp.~265-441.

\bibitem{kubo} R. Kubo, M. Toda, and N. Hashitsume, \textit{Statistical
Physics}, Vol.II (Springer-Verlag, Berlin, 1985).

\bibitem{lw} K. Lindenberg and B. J. West,
\textit{The Nonequilibrium Statistical Mechanics of Open and Closed
Systems} (VCH Publisher, Inc., New York, 1990).

\bibitem{skb} S.K. Banik, J. Ray Chaudhuri, and D.S. Ray, J. Chem.
Phys. \textbf{112}, 8330 (2000).

\bibitem{jrc} J. Ray Chaudhuri, S.K. Banik, B.C. Bag, and D.S.
Ray, Phys. Rev. E \textbf{63}, 061111 (2001).

\bibitem{motor} R.D. Astumian, Science {\bf 276}, 917 (1997);
P.Reimann, Phys. Rep. {\bf 361}, 57 (2002).

\bibitem{zwpf} R. Zwanzig, Phys. Fluid. \textbf{2}, 12 (1978).

\bibitem{cn} B. Carmeli and A. Nitzan, J. Chem. Phys. \textbf{79},
393 (1983).

\bibitem{rattray} K.M. Rattray and A.J. McKane, J. Phys. A \textbf{24},
4375 (1991).

\bibitem{nnds} See, for example, \textit{Noise in Nonlinear Dynamical Systems},
edited by F. Moss and P.V.E. McClintock (Cambridge University Press,
Cambridge, 1989), Vols.I-III.

\bibitem{masoliver} J. Masoliver and J.M. Porr\`a, Phys. Rev. E
\textbf{48}, 4309 (1993).

\bibitem{sjbe} S.J.B. Einchcomb and A.J. McKane, Phys. Rev. E
\textbf{49}, 259 (1994).

\bibitem{rig1}E. Hershkovits and R. Hernandez,
J. Chem. Phys. \textbf{122}, 014509 (2005).

\bibitem{lee} H. W. Hsia, N. Fang and X. Lee, Phys. Lett. A
\textbf{215}, 326 (1996); A. N. Drozdov and S. C. Tucker, J. Phys.
Chem. B \textbf{105}, 6675 (2001).

\bibitem{rig2} J.M. Moix and R. Hernandez, J. Chem. Phys. \textbf{122}, 114111 (2005).

\bibitem{nit} W. Horsthemke and R. Lefever, \textit{Noise-Induced
Transitions} (Springer-Verlag, Berlin, 1984).

\bibitem{dipole} L.D. Landau and E.M. Lifshitz,
\textit{The Classical Theory of Fields} (Pergamon, Oxford, 1975).

\bibitem{bravo} J.M. Bravo, R.M. Velasco, and J.M. Sancho, J.
Math. Phys. \textbf{30}, 2023 (1989).

\bibitem{resib} P. Resibois and M. dc Leener, \textit{Chemical Kinetic
Theory of Fluids} (Wiley-Interscience, NY, 1977).

\bibitem{lax} M. Lax, Rev. Mod. Phys. \textbf{38}, 541 (1966).

\bibitem{hw} P. H\"anggi and U. Weiss, Phys. Rev. A \textbf{29},
2265 (1984).

\bibitem{gh} R.F. Grote and J.T. Hynes, J. Chem. Phys. \textbf{77},
3736 (1982).

\bibitem{farkas} L. Farkas, Z. Phys. Chem. (Leipzig) \textbf{125},
236 (1927).

\bibitem{bhl} M. B\"uttiker, E.P. Harris, and R. Landauer, Phys.
Rev. B \textbf{28}, 1268 (1983).

\bibitem{alpha} M. B\"uttiker in
\textit{Noise in Nonlinear Dynamical Systems}, edited by F. Moss and
P.V.E. McClintock, Vol.~2 (Cambridge University Press, Cambridge,
1989) pp.~45-64.

\bibitem{heun1} T.C. Gard, in
\textit{Monographs and Textbooks in Pure and Applied Mathematics}
(Marcel Dekker, New York, 1987), Vol.114.

\bibitem{heun2} R. Toral, in \textit{Computational Field Theory and Pattern
Formation}, edited by P.L. Garrido and J. Marro, Lecture Notes in
Physics, Vol.448 (Springer-Verlag, Berlin, 1995).

\bibitem{mfpt} C. Mahanta and T.G. Venkatesh, Phys. Rev. E
\textbf{58}, 4141 (1998). J.M. Sancho, A.H. Romero, and K.
Lindenberg, J. Chem. Phys. \textbf{109}, 9888 (1998).

\bibitem{db} D. Barik, B.C. Bag and D.S. Ray, J. Chem. Phys.
\textbf{119}, 12973 (2003).

\bibitem{jrc2} J. Ray Chaudhuri, G. Gangopadhyay, and D.S. Ray,
J. Chem. Phys. \textbf{109}, 5565 (1998).

\bibitem{rig3} R. Hernandez, J. Chem. Phys. \textbf{111}, 7701 (1999).

\end{thebibliography}
\end{document}